\documentstyle[aps,preprint,amsmath,amssymb]{revtex}
\tighten
\begin{document}
\draft
\title{Renormalization Group Approach \\ to Cosmological Back Reaction 
Problems}
\date{May 23, 2000}
\author{Yasusada Nambu\footnote{e-mail:~nambu@allegro.phys.nagoya-u.ac.jp}}
\address{Department of Physics, Graduate School of Science, Nagoya University \\ 
        Chikusa, Nagoya 464-8602, Japan}
%
%
\preprint{DPNU-00-10}
\maketitle
\begin{abstract}
 We  investigated the  back reaction of cosmological perturbations on the evolution of the universe using the
second order perturbation of the Einstein's equation. To incorporate the
back reaction effect due to the inhomogeneity into the framework of the
cosmological perturbation, we used the renormalization group
method. The second order zero mode solution which appears by the
non-linearities of the Einstein's  equation is regarded as a secular term of the perturbative
expansion, we renormalized a constant of integration contained in
the background solution and absorbed the secular term to this
constant. For a dust dominated universe, using the second order gauge invariant quantity, we derived the renormalization group equation which determines the effective dynamics of the Friedman-Robertson-Walker universe with  the back reaction effect in a gauge invariant manner. We obtained the solution of the renormalization group equation and found that perturbations of the scalar mode and the long wavelength tensor mode works as  positive spatial curvature, and the short wavelength tensor mode  as  radiation fluid. 
\end{abstract}
\def\al{\alpha}
\def\del{\delta}
\def\pa{\partial}
\def\bk{\boldsymbol{k}}
\section{introduction}
Our universe seems to be very close to a Friedman-Robertson-Walker
(FRW) spacetime at length scale of the order of the Hubble radius, but
the metric and matter content appears to be highly inhomogeneous at
smaller scale. The conventional cosmological perturbation approach\cite{kodama,mukhanov2}
treat such a situation as the homogeneous isotropic background plus
small perturbation, and investigate the evolution of linear
fluctuations.  We must go beyond linear approximation to treat non-linear 
structure and construct suitable model of inhomogeneous universe
which is close to a FRW universe on large scales.

Based on the perturbative approach, the metric of inhomogeneous universe
is approximated by perturbative series and expanded as the background
FRW metric, the first order linear perturbation and the second order
perturbation which carries the non-linearities of the Einstein's equation. 
The non-linear effect affects not only the evolution of the  fluctuation, but
also change the evolution of  the background FRW universe.
This is the cosmological back reaction effect. Non-linear structure of
the universe can affect the dynamics of the background FRW universe and it is
important task to  understand  the effect of the
cosmological back reaction theoretically to determine the correct cosmological
parameter of our universe.

There are many works treating the back reaction problem so
far. Issacson\cite{issacson} considered the gravitational wave
propagating on a slowly varying background spacetime.  Assuming 
 high frequency of the gravitational wave, he expanded the vacuum 
Einstein's equation perturbatively. The leading order of the expansion
gives the propagation equation of the gravitational wave and the next
order gives the equation which determines the slowly varying
background spacetime on which the gravitational wave propagates. The
leading and the next order equations must be solved self-consistent
way.  Futamase\cite{futamase1,futamase2} discussed the effect of the
inhomogeneity on the global expansion rate of the universe using the
cosmological post-Newtonian approximation with specific gauge conditions.
Russ {\it et al.}\cite{russ2} treated the problem using the second
order perturbation with synchronous comoving gauge and derived the
evolution equation for the comoving volume. Boersma\cite{boersma} used
the linear perturbation with a uniform Hubble slice and derived the
spatially averaged Einstein's equation.
Abramo\cite{abramo1,abramo2,mukuhanov1,abramo3} derived the gauge
invariant effective energy momentum tensor for the scalar field and
the gravitational wave, and discussed the effect of inhomogeneity on
the background FRW universe in a gauge invariant manner.

These works are based on perturbation expansion of the Einstein's
equation with the averaging of the metric. The averaging operation of
the Einstein's equation is introduced to obtain the evolution equation
of the effective scale factor. Papers\cite{futamase2,russ2,boersma}
use the different gauge conditions and the different perturbation
scheme, and the relation of their result is not
obvious. Abramo\cite{abramo1,abramo2,mukuhanov1,abramo3} constructed
the gauge invariant effective energy momentum tensor and discussed the
effect of inhomogeneity on the background FRW universe, but he did not
derived explicit form of the evolution equation of the effective scale
factor.

In this paper, we investigate the cosmological back reaction problem
using the second order cosmological perturbation and the
renormalization group method\cite{chen,kunihiro,sasa,nozaki,nambu,boya1,boya2,ei}.
 We treat a dust dominated universe and examine how the inhomogeneity affect the 
 evolution of the background FRW universe.  
We does not use the averaging procedure for the Einstein's equation to obtain 
the evolution equation of the effective scale factor. We only use the
solution of the second order perturbation.  Within the framework of
the standard cosmological perturbation, we do not know how
to include the second order non-linear effect in the background FRW
metric because the perturbation of each order separate. We use the
method of the renormalization to incorporate the back reaction effect in
the perturbation expansion. Using this method, we can interpret the
back reaction as renormalization of constants of integration
contained in the background FRW solution. Furthermore, it is easy to
obtain the approximate analytic solution of the effective scale factor
which contains the back reaction effect of inhomogeneity. To circumvent the 
problem of gauge ambiguity, we construct the second order gauge invariant 
quantities and evaluate the effect of the back reaction in a gauge invariant 
manner. We find that the scalar mode and the long wavelength tensor mode perturbation
 work as positive spatial curvature and the short wavelength tensor mode perturbation
  works as radiation fluid.

The plan of this paper is as follows. In Sec. II, we introduce the
renormalization group method by using FRW model. In Sec. III, we
consider the second order cosmological perturbation of the Einstein's
equation and obtain the zero mode solution of the second order
perturbation. In Sec. IV, the second order gauge invariant quantity is
introduced. The renormalization group method is applied to the
solution of the second order zero mode perturbation and the effect of
the back reaction is discussed. Sec. V is devoted to summary and
discussion. We use the units in which $c=\hbar=8\pi G=1$ throughout the paper.
%

\section{renormalization group method}
In this section, we introduce the renormalization group
method\cite{chen,kunihiro,sasa,nozaki,nambu} using a FRW cosmological
model. To obtain temporal evolution of the solution of the non-linear 
differential equation, we usually apply a perturbative expansion. But
naive perturbation often yields secular terms due to resonance
phenomena. The secular term prevents us from getting approximate but
global solutions. The renormalization group method is one of
techniques to circumvent the problem. Starting from a naive
perturbative expansion, the secular divergence is absorbed into 
constants of integration contained in the zeroth order solution by the
renormalization procedure. The renormalized constants obey the
renormalization group equation.

We consider a dust dominated FRW universe with perfect fluid. The Einstein's equations are
\begin{subequations}
   \begin{align}
      & 
      \dot\al^{2}=\frac{1}{3}\,\rho_{0}+\frac{1}{3}\,\rho_{1},\label{eq:frw1}\\
      & \ddot\al+\frac{3}{2}\,\dot\al^{2}=-\frac{1}{2}\,p_{1},
   \end{align}
\end{subequations}
where $\al$ is logarithm of a scale factor of the universe,
$\rho_{0}$ is the energy density of the dust field, $\rho_{1}$ and
$p_{1}$ are the energy density and the pressure of perfect fluid with the
equation of the state $p_{1}=(\Gamma-1)\rho_{1}$ where $\Gamma$ is a
constant. Conservation equation for fluids are
\begin{subequations}
    \begin{align}
        &\dot\rho_{0}+3\,\dot\al\,\rho_{0}=0, \\
        &\dot\rho_{1}+3\,\Gamma\,\dot\al\,\rho_{1}=0,
    \end{align}
\end{subequations}
and their solutions are
\begin{equation}
    \rho_{0}=\frac{3\,c_{0}}{e^{3\al}},\quad 
    \rho_{1}=\frac{3\,c_{1}}{e^{3\Gamma\al}},
\end{equation}
where $c_{0}, c_{1}$ are constants of integration. Substitute this 
solution to Eq.(\ref{eq:frw1}), we obtain
\begin{equation}
    \dot\al^{2}=\frac{c_{0}}{e^{3\al}}+\frac{c_{1}}{e^{3\Gamma\al}} \label{eq:frwhc}
\end{equation}
We solve this equation perturbatively by assuming the second term of
the right hand side is small:
\begin{equation}
    \al=\al_{0}+\al_{1}+\cdots.
\end{equation}
The solution of the scale factor up to the first order of perturbation
is given by
\begin{equation}
      a(t)=a_{0}\,t^{2/3}\left[1+\frac{3\,c_{1}}{4\,(3-2\Gamma)}\,a_{0}^{-3\Gamma}
       (t^{2-2\Gamma}-t_{0}^{2-2\Gamma})\right], \label{eq:frnaive}
\end{equation}
where $a_{0}$ and $t_{0}$ are constants of integration of the zeroth
order and the first order, respectively. For $0<\Gamma<1$, the first
order term grows in time and the perturbation breaks down for large
$t$. We improve this behavior by applying the renormalization group
method\cite{chen,nozaki,nambu}.

We redefine the zeroth order integration constant $a_{0}$ as
\begin{equation}
    a_{0}=a_{0}^{R}(\mu)+\del a_{0}(t_{0};\mu),
\end{equation}
where $\mu$ is a renormalization point and $\del a_0$ is a counter
term which absorbs the secular divergence of the solution. The naive solution
(\ref{eq:frnaive}) can be written
\begin{align}
   a(t)&= t^{2/3}\left[a_{0}^{R}(\mu)+\del a_{0}^{R}(t_{0};\mu)
         +\frac{3\,c_{1}\,(a_{0}^{R})^{1-3\Gamma}}{4\,(3-2\Gamma)}      
 (t^{2-2\Gamma}-\mu^{2-2\Gamma}+\mu^{2-2\Gamma}-t_{0}^{2-2\Gamma})
 \right] \notag\\
      &=t^{2/3}\left[a_{0}^{R}(\mu)
         +\frac{3\,c_{1}\,(a_{0}^{R})^{1-3\Gamma}}{4\,(3-2\Gamma)}      
 (t^{2-2\Gamma}-\mu^{2-2\Gamma})
 \right], \label{eq:frwnaive2}
\end{align}
where we have chosen the counter term $\del a_{0}$ so as to absorb the
$(\mu^{2-2\Gamma}-t_0^{2-2\Gamma})$ dependent term:
\begin{equation}
    \del a_{0}(t_{0};\mu)=a_{0}^{R}(t_{0})-a_{0}^{R}(\mu)=
     -\frac{3\,c_{1}\,(a_{0}^{R})^{1-3\Gamma}}{4\,(3-2\Gamma)}
     (\mu^{2-2\Gamma}-t_{0}^{2-2\Gamma}). \label{eq:counter}
\end{equation}
This defines the renormalization transformation
\begin{equation}
    {\mathcal R}_{\mu, t_{0}}~:~a_{0}^{R}(t_{0})\longmapsto
     a_{0}^{R}(\mu)=a_{0}^{R}(t_{0})-\frac{3\,c_{1}\,(a_{0}^{R})^{1-3\Gamma}}{4\,(3-2\Gamma)}
     (\mu^{2-2\Gamma}-t_{0}^{2-2\Gamma}),
\end{equation}
and this transformation forms the Lie group up to the first order of
the perturbation. So we can have $a_0(\mu)$ for arbitrary large value
of $(\mu^{2-2\Gamma}-t_0^{2-2\Gamma})$ by assuming the property of the
Lie group, and this makes it possible to produce a globally uniform
approximated solution of the original equation. 
The renormalization group equation is obtained by differentiating 
Eq.(\ref{eq:counter}) with respect to $\mu$, and setting $t_{0}=\mu$:
\begin{equation}    
    \frac{\pa}{\pa\mu}a_{0}^{R}(\mu)=\frac{3\,c_{1}(1-\Gamma)}
      {2\,(3-2\Gamma)}\, a_{0}^{1-3\Gamma}\mu^{1-2\Gamma}.
\end{equation}
The renormalized solution is obtained by equating $\mu=t$ in 
(\ref{eq:frwnaive2}):
\begin{equation}
   a^{R}(t)=t^{2/3}\,a_{0}^{R}(t)=t^{2/3}\left[c+\frac{9\,c_{1}\,\Gamma}
     {4\,(3-2\Gamma)}\,t^{2-2\Gamma}\right]^{1/(3\Gamma)}.
\end{equation}

Now we compare the renormalized solution with the naive solution for 
$\Gamma=2/3$ and $\Gamma=4/3$ cases. For $\Gamma=2/3$, the energy
density of the fluid is $\rho_1\propto a^{-2}$ and the effect of the 
fluid is equivalent to  spatial curvature. The naive and the renormalized 
solutions are
\begin{subequations}
\begin{align}
    a(t)&=a_{0}\,t^{2/3}\left[1+\frac{9\,c_{1}}{20\,a_{0}^{2}}\,(t^{2/3}
       -t_{0}^{2/3})\right], \\
    a^{R}(t)&=t^{2/3}\left(c+\frac{9\,c_{1}}{10}\,t^{2/3}\right)^{1/2}.
\end{align}
\end{subequations}
To see  how the renormalized solution improves the naive 
solution, we choose $a_0=1, c_1=-1, c=1, t_0=0$. $c_{1}<0$ corresponds
to the positive spatial curvature. In this case, the exact solution of
Eq.(\ref{eq:frwhc}) is given by
\begin{equation}
\begin{split}
  &a=\frac{9}{2}\,(1-\cos\eta), \\
  &t=\frac{9}{2}\,(\eta-\sin\eta)\qquad (0\le\eta\le 2\pi).
\end{split}
\end{equation}
The scale factor of the exact solution has a maximum value $4/9\sim
0.4444$ at $t=2\pi/9\sim 0.6981$. The naive solution has a maximum
value $5/9\sim 0.5556$ at $t=\left(10/9\right)^{3/2}\sim 1.1712$. The
renormalized solution has a maximum value $20/(27\sqrt{3})\sim 0.4277$
at $t=(20/27)^{3/2}\sim 0.6375$.  The renormalized solution gives
accurate approximated solution and reproduces the contracting behavior
of the universe qualitatively and quantitatively well.

For $\Gamma=4/3$, the energy density of the fluid is $\rho_1\propto
a^{-4}$ and the effect of the fluid is the same as radiation. 
The naive and the renormalized solutions are
\begin{subequations}
    \begin{align}       
        a(t)&=a_{0}\,t^{2/3}\left[1+\frac{9\,c_{1}}{4\,a_{0}^{4}}\,(t^{-2/3}
           -t_{0}^{-2/3})\right],\label{eq:radnaive} \\
        a^{R}(t)&=t^{2/3}\left(c+9\,c_{1}\,t^{-2/3}\right)^{1/4}
    \end{align}
\end{subequations}
The renormalized solution behaves as $a^{R}\propto t^{1/2}$ for $t\sim 0$ 
and $a^{R}\propto t^{2/3}$ for $t\sim\infty$. Therefore the 
renormalized solution reproduces the behavior of the scale factor
which represents the transition from the radiation 
dominant era to the matter dominant era. We cannot get such a
behavior from the naive solution (\ref{eq:radnaive}).

\section{zero mode solution of cosmological second order perturbation}
We aim to treat the cosmological back reaction problem using
perturbation approach. Let us assume that the metric is expanded as follows:
\begin{equation}
  g_{ab}=\overset{(0)}g_{ab}+\overset{(1)}g_{ab}+\overset{(2)}g_{ab}+\cdots.
\end{equation}
$\overset{(0)}g_{ab}$ is the background FRW metric and represent the
homogeneous and isotropic space. $\overset{(1)}g_{ab}$ is the metric
of the first order (linear) perturbation. This metric represents the
small linear fluctuation from the background space time. We can assume
that the spatial average of the first order perturbation vanishes:
\begin{equation}
  \langle\overset{(1)}g_{ab}\rangle=0,
\end{equation}
where $\langle\cdots\rangle$ means the spatial average with respect to
the background FRW metric. $\overset{(2)}g_{ab}$ is the second order
metric and contains non-linear effect caused by the first order linear
perturbation. This non-linearity  produces homogeneous and isotropic zero mode
part of the second order metric. That is
\begin{equation}
  \langle\overset{(2)}g_{ab}\rangle \ne 0.
\end{equation}
As we want to interpret the zero mode part of the metric as the background FRW metric,  we must redefine the background metric as follows:
\begin{equation}
  \overset{(0)}g_{ab}~\longrightarrow \overset{(0)}g_{ab}+
     \langle\overset{(2)}g_{ab}\rangle. \label{eq:redef}
\end{equation}
This is the back reaction caused by the non-linearities  of the
fluctuation and it changes the background metric. But in cosmological situations, the
second order perturbation term grows in time and dominates the
background metric, this simple prescription does not work well in
the context of the perturbation expansion. Furthermore, the meaning of
the gauge invariance is not obvious in the second order quantity. 
We cannot adopt Eq.(\ref{eq:redef}) as the definition of the
background metric because the gauge transformation changes the
definition of the background metric. We will resolve these problems in the
next section by constructing the second order gauge invariant quantity and applying the renormalization group method.

In this section, we obtain the solution of the second order
perturbation.  To extract the effect of the back reaction, it is not
necessary to know the full form of the solution. We only need the
homogeneous and isotropic zero mode of the second order perturbation. We
consider the perturbation of the Einstein's equation with dust field in
comoving synchronous gauge. The background is assumed to be a spatially flat
FRW universe. The metric is
\begin{equation}
  ds^2=-dt^2+g_{ij}(t,\boldsymbol{x})dx^idx^j.
\end{equation}
The Einstein's equations are
\begin{subequations}
  \begin{align}
    &{}^{(3)}R+K^2-K^i{}_jK^j{}_i=2\,\rho, \\
    & K_i{}^j{}_{|j}-K_{|i}=0, \\
    & \dot K^i{}_j+KK^i{}_j+{}^{(3)}R^i{}_j=\frac{\rho}{2}\,\del^i{}_j.\,
  \end{align}
\end{subequations}
where ${}^{(3)}R_{ij}$ is the Ricci tensor of the spatial metric, $\rho$ is the energy
density of the dust and the extrinsic curvature is defined by
\begin{equation}
  K^j{}_i=\frac{1}{2}\,g^{jk}\,\dot g_{ki}.
\end{equation}
By eliminating $\rho$, the evolution equation can be written
\begin{equation}
  \dot K^j{}_i+KK^j{}_i-\frac{1}{4}\,\del^j{}_i(K^2-K^l{}_mK^m{}_l)
    +\left({}^{(3)}R^j{}_i-\frac{1}{4}\,\del^j{}_i{}^{(3)}R\right)=0.
\end{equation}
Using the conformally transformed metric
$\gamma_{ij}=a^{-2}(t)\,g_{ij}$,
\begin{equation}
  K^j{}_i=H\,\del^j{}_i+\frac{1}{2}\,\gamma^{jl}\,\dot\gamma_{li},
\end{equation}
where $H=\dot a/a$ and $a$ is the scale factor of the background FRW universe. 
The  evolution equation for the three metric becomes
\begin{equation}
  \label{eq:evo3}
  \begin{split}
  \dot
  k^j{}_i&+3H\,k^j{}_i+\frac{2}{a^2}\left(R^j{}_i-\frac{1}{4}\del^j{}_i
    R\right)
  \\  
   &=-\del^j{}_i\,(2\dot H+3H^2)-\frac{1}{2}\,k\,k^j{}_i+\frac{1}{8}\,\del^j{}_i
    \,(k^2-k^l{}_m\,k^m{}_l),
  \end{split}
\end{equation}
where $R_{ij}$ is the Ricci tensor of the conformally transformed
metric $\gamma_{ij}$ and $k_{ij}=\dot\gamma_{ij}$. We solve this
equation perturbatively up to the second order:
\begin{equation}
  \gamma_{ij}=\del_{ij}+h_{ij}+l_{ij},
\end{equation}
where $h_{ij}$ is the metric of the first order and $l_{ij}$ is the metric of
the second order, respectively.

\subsection{Background solution}
The background evolution equation  is
\begin{equation}
  2\dot H+3H^2=0,
\end{equation}
and the solution is
\begin{equation}
    a(t)=a_{0}\,t^{2/3},
\end{equation}
where $a_0$ is a constant of integration. 

\subsection{First order solution}
The first order evolution equation is
\begin{equation}
  \ddot h^j{}_i+3H\,\dot h^j{}_i+\frac{2}{a^2}\left(
    \overset{(1)}R{}^j{}_i(h)-\frac{1}{4}\del^j{}_i\overset{(1)}R(h)\right)=0,
\end{equation}
where $\overset{(1)}R_{ij}(h)$ is the linear part of the Ricci
tensor:
\begin{equation}
  \overset{(1)}R_{ij}(h)=\frac{1}{2}\,(-h_{,ij}-\nabla^2h_{ij}+h_{ki,j}{}^{,k}
      +h_{kj,i}{}^{,k}).
\end{equation}
 The momentum constraint is
\begin{equation}
  h_i{}^j{}_{,j}-h_{,i}=0. \label{eq:mc}
\end{equation}
\subsubsection{scalar mode}
For the scalar mode perturbation, the metric can be written using
scalar functions as follows:
\begin{equation}
  h_{ij}^{(S)}=A(t,\boldsymbol{x})\,\del_{ij}+B_{,ij}(t,\boldsymbol{x}).
\end{equation}
By using the momentum constraint (\ref{eq:mc}), we have
\begin{equation}
  \dot A(t,\boldsymbol{x})=0.
\end{equation}
Therefore
\begin{equation}
  A(t,\boldsymbol{x})=\frac{20}{9}\,\Psi(\boldsymbol{x}),
\end{equation}
where $\Psi$ is an arbitrary function of the spatial coordinate. The
evolution equation becomes
\begin{equation}
  (\nabla^2B)\,\ddot{}+3H(\nabla^2
  B)\,\dot{}-\frac{20}{27\,a^2}\nabla^2\Psi =0,
\end{equation}
and the growing mode solution is
\begin{equation}
  B=\frac{2\,t^{2/3}}{a_0^2}\,\Psi.
\end{equation}
The scalar mode solution is given by
\begin{equation}
  \label{eq:scalarsol}
  h_{ij}^{(S)}=\frac{20}{9}\,\Psi(\boldsymbol{x})\,\del_{ij}
      +\frac{2\,t^{2/3}}{a_0^2}\,\Psi_{,ij}(\boldsymbol{x}).
\end{equation}
The density contrast is given by
\begin{equation}
  \label{eq:density}
  \frac{\overset{(1)}\rho}{\overset{(0)}\rho}=-\frac{4}{9\,a^2H^2}\,\nabla^2\Psi
   =-\frac{t^{2/3}}{a_0^2}\,\nabla^2\Psi\propto a(t).
\end{equation}

\subsubsection{tensor mode}
The evolution equation for the tensor mode is
\begin{equation}
  \ddot h_{ij}^{(GW)}+3H\,\dot h_{ij}^{(GW)}-\frac{1}{a^2}\,\nabla^2
  h_{ij}^{(GW)}=0,
\end{equation}
where $h_{ij}^{(GW)}$ satisfies the transverse traceless condition:
\begin{equation}
  h_i{}^j{}_{,j}=h_i{}^i=0.
\end{equation}
The solution is given by
\begin{equation}
  \label{eq:tensorsol}
  \begin{split}
   &h_{ij}^{(GW)}=\sum_{\bk}e^{i\bk\cdot\boldsymbol{x}}\,A_{ij}(t,\bk), \\
   & ; A_{ij}(t,\bk)=A_{ij}(\bk)\,t^{-1/2}\,
    Z_{\pm 3/2}\left(\frac{3\,k}{a_0}\,t^{1/3}\right),    
  \end{split}
\end{equation}
where $Z$ is a Bessel function and $A_{ij}(\bk)$ is a constant tensor
satisfying the transverse traceless condition.

\subsection{Second order solution}
The evolution equation for the second order is
\begin{equation}
  \begin{split}
  \ddot l^j{}_i+3H\,\dot l^j{}_i&+\frac{2}{a^2}
    \left(\overset{(2)}R{}^{Lj}{}_i-\frac{1}{4}\,\del^j{}_i\overset{(2)}R
      {}^L\right) = \\
 &-\frac{2}{a^2}\,h^{jk}\left(\overset{(1)}R{}_{ki}-\frac{1}{4}\,\del_{ki}
    \overset{(1)}R\right)+\dot h^{jk}\,\dot h_{ki}-\frac{1}{2}\,\dot h
     \,\dot h^j{}_i+\frac{1}{8}\,\del^j{}_i\,(\dot h^2-\dot h^k{}_l\,\dot
     h^l{}_k)\\
 &-\frac{2}{a^2}\left[\overset{(2)}R{}^{NLj}{}_i-h^{jk}\,\overset{(1)}R_{ki}
   -\frac{1}{4}\del^j{}_i\left(\overset{(2)}R{}^{NL}-h^{kl}\,\overset{(1)}R_{kl}\right)\right],
  \end{split}
  \label{eq:evo2}
\end{equation}
where $\overset{(2)}R{}^L_{ij}$ is the linear part of the Ricci tensor
with respect to the second order metric $l_{ij}$ and
$\overset{(2)}R{}^{NL}_{ij}$ is the quadratic part of the Ricci tensor
with respect to the first order metric $h_{ij}$. Due to the
homogeneity and the isotropy of the background metric, the zero mode
part of the second order metric satisfies
\begin{equation}
  \langle l_{ij}\rangle =\frac{1}{3}\,\del_{ij}\,\langle l\rangle,
\end{equation}
where the spatial average is defined as
\begin{equation}
 \langle l_{ij}\rangle=\frac{1}{V}\int_V d^3x\,l_{ij}.
\end{equation}
$V$ is the volume of a sufficiently large compact domain and we assume periodic boundary conditions for perturbations. 
Therefore it is sufficient to consider only the trace part of the second
order metric.  The trace part of the evolution equation
(\ref{eq:evo2}) with the spatial average becomes
\begin{equation}
 \label{eq:evo3}
  \langle l\rangle\,\ddot{}+3H\,\langle l\rangle\,\dot {}=
 \Bigl\langle\frac{5}{8}\,\dot h^{kl}\,\dot h_{kl}-\frac{1}{8}\,\dot
 h^2-\frac{2}{a^2}
 \left(\frac{1}{4}\,\overset{(2)}R{}^{NL}-\frac{1}{4}\,h\,\overset{(1)}R
 +\frac{3}{4}\,h^{kl}\overset{(1)}R_{kl}\right)\Bigr\rangle,
\end{equation}
where
\begin{equation}
  \begin{split}
    \overset{(2)}R{}^{NL}=&\frac{1}{2}\,\Bigl[\frac{1}{2}\,h_{kl,i}\,h^{kl,i}+ 
     h^{kl}\,(\nabla^2h_{kl}+h_{,kl}-2\,h_{ki,l}{}^{,i}) \\
     &+h^{ik,l}\,(h_{ik,l}-h_{il,k})-
      \left(h^{kl}{}_{,l}-\frac{1}{2}\,h^{,k}\right)(2\,h_{ki}{}^{,i}-h_{,k})
      \Bigl].
  \end{split}
\end{equation}

\subsubsection{second order zero mode solution : contribution of scalar mode}

For the first order scalar mode solution (\ref{eq:scalarsol}),
Eq.(\ref{eq:evo3}) becomes
\begin{equation}
  \langle l\rangle \,\ddot{}+3H\,\langle l\rangle\,\dot{}=
   \frac{100}{81}\frac{t^{-4/3}}{a_0^2}\sum_{\bk}k^2\,\Psi_{\bk}\Psi_{\bk}^{*}
   +\frac{28}{9}\frac{t^{-2/3}}{a_0^4}\sum_{\bk}k^4\,\Psi_{\bk}\Psi_{\bk}^{*},
\end{equation}
where $\Psi_{\bk}$ is the Fourier component of $\Psi(\boldsymbol{x})$. 
The solution of this equation is
\begin{equation}
  \langle l\rangle
  =\frac{10}{9}\frac{t^{2/3}}{a_0^2}\sum_{\bk}k^2\,\Psi_{\bk}
   \Psi_{\bk}^{*}+\frac{t^{4/3}}{a_0^4}\sum_{\bk}k^4\,\Psi_{\bk}
   \Psi_{\bk}^{*}+\text{const.}.
\end{equation}
This solution is consistent with the result of Tomita and
Kasai\cite{tomita,kasai} who obtained the complete form of the second
order solution.

\subsubsection{second order zero mode solution : contribution of
  tensor mode}

For the first order tensor mode solution (\ref{eq:tensorsol}),
Eq.(\ref{eq:evo3}) becomes
\begin{equation}
 \label{eq:evo4}
 \langle l\rangle \,\ddot{}+3H\,\langle l\rangle\,\dot{}=\sum_{\bk}
 \left(\frac{5}{8}\,\dot A^{kl}(t,\bk)\dot A^{*}_{kl}(t,\bk)-
   \frac{7\,k^2}{8\,a^2}\,A^{kl}(t,\bk)A^{*}_{kl}(t,\bk)\right).
\end{equation}

For the mode whose wavelength is smaller than the horizon scale $k\gg
aH$, the first order solution is approximated to be WKB form:
\begin{equation}
  A_{ij}(t,\bk)\approx\frac{1}{a}\,A_{ij}(\bk)\,e^{ik\int\frac{dt}{a}},
  \quad A^i{}_i=k^i \,A_{ij}=0. 
\end{equation}
For this solution Eq.(\ref{eq:evo4}) becomes
\begin{equation}
  \langle l\rangle\,\ddot{}+3H\,\langle l\rangle\,\dot{}\approx 
   -\frac{1}{4\,a^4}\sum_{\bk}k^2\,A^{ij}(\bk)A^{*}_{ij}(\bk),
\end{equation}
and the solution is given by
\begin{equation}
  \langle l\rangle =\frac{9}{8}\frac{t^{-2/3}}{a_0^4}\sum_{\bk}
    k^2\,A^{ij}(\bk)A^{*}_{ij}(\bk).
\end{equation}

For the long wavelength mode whose wavelength is longer than the
horizon scale $k\ll aH$, the first order tensor mode solution becomes
\begin{equation}
  A_{ij}(t,\bk)\approx
  A_{ij}(\bk)\left[1-\frac{2}{5}\left(\frac{k}{aH}\right)^2\right],
\end{equation}
and the Eq.(\ref{eq:evo4}) becomes
\begin{equation}
  \langle l\rangle\,\ddot{}+3H\,\langle l\rangle\,\dot{}\approx
    -\frac{7}{8\,a^2}\sum_{\bk}k^2\,A^{ij}(\bk)A^{*}_{ij}(\bk).
\end{equation}
The solution is
\begin{equation}
  \langle l\rangle =-\frac{63}{80}\,\frac{t^{2/3}}{a_0^2}\sum_{\bk}
    k^2\,A^{ij}(\bk)A^{*}_{ij}(\bk).
\end{equation}
\section{gauge invariance and renormalization of the second order zero
  mode solution } 

The second order perturbation has the zero mode part which
comes from the the non-linearity of the Einstein's equation, and
this changes the evolution of the background FRW universe. We want to
treat this back reaction effect by using the renormalization
method. The background FRW solution has a constant of integration
which corresponds to the freedom of constant rescaling of the scale
factor. By absorbing the second order zero mode part to this constant
using the renormalization method, we can obtain the effective scale
factor of the FRW universe which contains the effect of back reaction
due to the inhomogeneity. But the existence of the gauge freedom
prevents us from direct application of the renormalization method. If
we change the gauge condition, the component of the second order zero
mode solution may change. We do not know what component or combination
of the zero mode solution should be renormalized.

In this section, we first review the method of construction of the
second order gauge invariant quantities using Abramo's
argument\cite{abramo1,abramo2,mukuhanov1}. Then we apply the
renormalization group method to the second order gauge
invariant quantity for the zero mode.  Let us consider the infinitesimal coordinate
transformation
\begin{equation}
  x^\mu \longrightarrow x'{}^\mu=x^\mu+\xi^\mu,
\end{equation}
where $\xi^\mu$ is assumed to be the first order small quantity. 
An arbitrary tensor $q$ receives the gauge transformation
\begin{equation}
  q\longrightarrow q'=e^{-{\mathcal L}_{\boldsymbol{\xi}}}\,q=q-
    {\mathcal L}_{\boldsymbol{\xi}}\,q+\frac{1}{2}{\mathcal
      L}^2_{\boldsymbol{\xi}}\,q
    +\cdots,
\end{equation}
where ${\cal L}_{\boldsymbol{\xi}}$ is the Lie derivative with respect
to the vector $\boldsymbol{\xi}$. 
Now we prepare the vector $\boldsymbol{X}$ which transforms under the
coordinate transformation as follows:
\begin{equation}
  \label{eq:xtransf}
  X^\mu \longrightarrow X' {}^\mu=X^\mu+\xi^\mu+\frac{1}{2}\,
   [ \boldsymbol{X},\boldsymbol{\xi}]^\mu+\cdots.
\end{equation}
Then the quantity $Q\equiv e^{{\mathcal L}_{\boldsymbol{X}}}\,q$
transforms as
\begin{align}
  Q \longrightarrow Q'&=e^{{\mathcal L}_{\boldsymbol{X}'}}\,q'=
      e^{{\mathcal
          L}_{\boldsymbol{X}'}}e^{-{\mathcal L}_{\boldsymbol{\xi}}}\,q \notag \\
    &=e^{{\mathcal L}_{\boldsymbol{X}}}\,q=Q.
\end{align}
Therefore the quantity $Q$ is gauge invariant. If expand the tensor
$q$ perturbatively
$q=\overset{(0)}{q}+\overset{(1)}{q}+\overset{(2)}{q}$, the gauge
invariant quantities in each order are
\begin{subequations}
\begin{align}
   \overset{(0)}Q&=\overset{(0)}q, \\
   \overset{(1)}Q&=\overset{(1)}{q}+{\mathcal
    L}_{\boldsymbol{X}}\overset{(0)}{q}, \\
   \langle\overset{(2)}Q\rangle&=\langle\overset{(2)}{q}\rangle+
     \langle{\mathcal L}_{\boldsymbol{X}}\overset{(1)}q\rangle+
     \frac{1}{2}\langle{\mathcal
       L}^2_{\boldsymbol{X}}\overset{(0)}{q}\rangle 
      +\langle{\mathcal L}_{\boldsymbol{Y}}\overset{(0)}q\rangle.
\end{align}
\end{subequations}
$\overset{(0)}Q$ is the background variable and does not receive gauge
transformation.  $\boldsymbol{Y}$ represents the vector which is
needed to construct the gauge invariant combination under the 
second order coordinate transformation. We does not need explicit form
of this vector. For the second order zero mode metric, the effect of
$\boldsymbol{Y}$ is to produce the second order time coordinate
transformation. We can set $\boldsymbol{Y}=0$ if we do not mind the
freedom of the second order time coordinate transformation. The
quantity $\langle\overset{(2)}Q\rangle $ is invariant under  first
order gauge transformation for $\boldsymbol{Y}=0$.

Now we look for the explicit form of the gauge invariant quantity in
a  specific gauge. We consider the following form of the first order
metric:
\begin{equation}
  \overset{(1)}g_{\mu\nu}=
  \begin{pmatrix}
   -2\,\phi & a\,B_{,i} \\
    a\,B_{,i} & a^2\,(-2\,\psi\,\del_{ij}+2\,E_{,ij})
  \end{pmatrix}.
\end{equation}
Under the coordinate transformation, the 
perturbation variables transform as follows
\begin{subequations}
  \begin{align}
    &\phi'=\phi-\dot\xi^0, \\
    &B'=B-a{\left(\frac{1}{a^2}\,\xi\right)}^{\centerdot}+\frac{1}{a}\,\xi^0, \\
    &\psi'=\psi+H\,\xi^0, \\
    &E'=E-\frac{1}{a^2}\,\xi,\\
    &\del\chi'=\del\chi-\xi^0\,\dot\chi_0,
  \end{align}
\end{subequations}
where the scalar function $\chi$ represents the velocity potential for
the dust field.  From this, we can choose the following  vector
$\boldsymbol{X}$ which obeys the desired transformation (\ref{eq:xtransf}):
\begin{equation}
     \boldsymbol{X}=[-\frac{\del\chi}{\dot\chi_0}, -\del_{ij}\,E_{,j}].
\end{equation}
Of course, the vector $\boldsymbol{X}$ is not unique. There are infinite
possibility of the form of $\boldsymbol{X}$ which corresponds to the
infinite possible candidate of the gauge invariant combination of the
perturbation variables. We choose this form because we want to impose
the comoving gauge condition $\del\chi=0$ for the perturbation
variables.  The first order gauge invariant quantities constructed from
$\boldsymbol{X}$ are
\begin{subequations}
  \begin{align}
    \overset{(1)}Q_{00}&=-2\,\phi+2{\left(\frac{\del\chi}{\dot\chi_0}\right)}^{\centerdot} ,\\
    \overset{(1)}Q_{0i}&= \left(a\,B-a^2\,\dot
    E+\frac{\del\chi}{\dot\chi_0}\right)_{,i}, \\
    \overset{(1)}Q_{ij}&=-2\,a^2\,\del_{ij}\left(
    \psi+\frac{H}{\dot\chi_0}\,\del\chi\right),\\
    \overset{(1)}Q_{\chi}&=\del\chi+X^0\,\dot\chi_0=0.
  \end{align}
\end{subequations}
For comoving synchronous gauge $\phi=B=\del\chi=0$, there is no
freedom of the coordinate transformation and the each component of the
metric perturbation can be written using the combination of these
gauge invariant variables.

\subsection{gauge invariant zero mode quantity: contribution of scalar mode}

Using the first order scalar mode solution (\ref{eq:scalarsol}), the zero mode
gauge invariant quantities in comoving synchronous gauge are
 
\begin{align}
& \langle\overset{(2)}Q_{00}\rangle=\langle\overset{(2)}Q_{0i}\rangle 
 =\langle\overset{(2)}Q_\chi\rangle=0, \notag \\
 & \overset{(0)}Q_{ij}+\langle\overset{(2)}Q_{ij}\rangle=a^2\,\del_{ij}+\langle\overset{(2)}{g}_{ij}\rangle
       +\langle{\mathcal
         L}_{\boldsymbol{X}}\overset{(1)}{g}_{ij}\rangle
       +\frac{1}{2}\,\langle{\mathcal
         L}^2_{\boldsymbol{X}}\overset{(0)}{g}_{ij} \rangle \notag \\
          &\qquad\qquad  =a^2\,\del_{ij}\,\left[
            1+\frac{2}{3}\,\langle\psi_{,k}E_{,k}\rangle
             -\frac{1}{3}\,\langle E_{,kl}E_{,kl}\rangle\right]
           +\langle\overset{(2)}{g}_{ij}\rangle \notag \\
         &\qquad\qquad  =a^2\left[1-\frac{10\,t^{2/3}}{27\,a_0^2}\sum_{\bk}k^2\,\Psi_{\bk}\Psi_{\bk}\right] \del_{ij}.
\end{align}
These quantities are invariant under the first order gauge
transformation. To interpret the geometrical meaning of the gauge
invariant quantity
$\overset{(0)}Q_{ij}+\langle\overset{(2)}Q_{ij}\rangle$, we consider
the determinant of the three metric:
\begin{align}
  \langle\sqrt{g}\rangle &=
   a^3\left[1+\langle\psi_{,i}\,E_{,i}\rangle-\frac{1}{2}\,\langle
     E_{,ij}\,E_{,ij}\rangle+\frac{3}{2}\,\langle\psi^2\rangle+\frac{1}{2}\,\langle\overset{(2)}g\rangle\right]\notag \\
 &=\left[\frac{1}{3}\,\del^{ij}\,(\overset{(0)}Q_{ij}+\langle\overset{(2)}Q_{ij}\rangle)+a^2\,(\langle\psi^2\rangle+\text{const.})\right]^{3/2} \notag \\
 &=\left[\frac{1}{3}\,\del^{ij}\,(\overset{(0)}Q_{ij}+\langle\overset{(2)}Q_{ij}\rangle)\right]^{3/2}.
\end{align}
Therefore, the trace of the gauge invariant variable
$\overset{(0)}Q_{ij}+\langle\overset{(2)}Q_{ij}\rangle$ corresponds to
the physical volume element of the universe and we can interpret
$\overset{(0)}Q+\langle\overset{(2)}Q\rangle$ as square of the scale factor of
the FRW universe:
\begin{equation}
  \label{eq:frwnaive}
  \begin{split}
   &ds^2=-dt^2+\tilde a^2(t)\,d\boldsymbol{x}^2, \\
   &\quad ;\quad \tilde a(t)=\left(\overset{(0)}Q+\langle\overset{(2)}Q\rangle\right)^{1/2}
     =t^{2/3}\left(a_0-\frac{5\,t^{2/3}}{27\,a_0}\sum_{\bk}k^2\,
       \Psi_{\bk}\Psi^{*}_{\bk}\right).
  \end{split}
\end{equation}
At this stage, we renormalize the expression for the effective scale
factor $\tilde a(t)$ to improve its behavior for large $t$. The background FRW solution has the constant of integration $a_0$. We regard the second order term in $\tilde a(t)$ as a secular term and apply the renormalization group method. The
renormalization group equation for $a_0$ becomes
\begin{equation}
  \frac{\pa a_0}{\pa t^{2/3}}=-\frac{5}{27\,a_0}\sum_{\bk}k^2\,
     \Psi_{\bk}\Psi^{*}_{\bk},
\end{equation}
and the solution is
\begin{equation}
  a_0(t)=\left(c-\frac{10\,t^{2/3}}{27}\sum_{\bk}k^2\,
       \Psi_{\bk}\Psi^{*}_{\bk}\right)^{1/2},
\end{equation}
where $c$ is a constant of integration. 
The renormalized expression of the line element (\ref{eq:frwnaive}) is
\begin{equation}
  \begin{split}
  &ds^2=-dt^2+(a^R(t))^2\,d\boldsymbol{x}^2, \\
  &\quad ; a^R(t)=t^{2/3}\left(c-\frac{10\,t^{2/3}}{27}
               \sum_{\bk} k^2\,\Psi_{\bk}\Psi^{*}_{\bk}\right)^{1/2}.
  \end{split}
\end{equation}
Comparing with the analysis of the section II, this solution is the
same as the FRW equation with  positive spatial curvature. We
conclude that the effect of the inhomogeneity due to the scalar mode
fluctuation is equivalent to  positive spatial curvature. This
result is the same as Russ {\it et al.}\cite{russ2} who derived the
spatially averaged Einstein's equation for a comoving volume and evaluate its behavior using the solution of the second
order perturbation of the Einstein's equation. 

\subsection{gauge invariant zero mode quantity: contribution of tensor mode}

For the first order tensor mode perturbation, the second order
solution $\langle l_{ij}\rangle$ is invariant under the first order
gauge transformation because the first
order tensor mode is already gauge invariant. The gauge invariant zero mode 
quantity is given by
\begin{equation}
  \overset{(0)}Q_{ij}+\langle\overset{(2)}Q_{ij}\rangle=a^2\,(\del_{ij}+\langle l_{ij}\rangle).
\end{equation}
 Using the second order zero mode solution, we have
\begin{equation}
  \overset{(0)}Q_{ij}+\langle\overset{(2)}Q_{ij}\rangle=
    \begin{cases}
     &\displaystyle{ a^2\left(1+\frac{3}{8}\frac{t^{-2/3}}{a_0^4}\sum_{\bk}
           k^2A^{ij}(\bk)A^{*}_{ij}(\bk)\right)\del_{ij}}\qquad
       \text{for}\quad k\gg aH\\
     &\displaystyle{
       a^2\left(1-\frac{21}{80}\frac{t^{2/3}}{a_0^2}\sum_{\bk}
          k^2 A^{ij}(\bk)A^{*}_{ij}(\bk)\right)\del_{ij}
                  } \qquad
       \text{for}\quad k\ll aH
    \end{cases}
\end{equation}
The renormalized scale factor becomes
\begin{equation}
  a^R(t)=
    \begin{cases}
      & \displaystyle{
         t^{2/3}\left(c+\frac{3}{4}\,t^{-2/3}\sum_{\bk}k^2
           A^{ij}(\bk)A^{*}_{ij}(\bk)\right)^{1/4}
         }\qquad \text{for}\quad k\gg aH \\
      & \displaystyle{
          t^{2/3}\left(c-\frac{21}{80}\,t^{2/3}\sum_{\bk}k^2
            A^{ij}(\bk)A^{*}_{ij}(\bk)\right)^{1/2}
          } \qquad \text{for}\quad k\ll aH
    \end{cases}
\end{equation}
Comparing with the result of section II, the effect of the back reaction
of the short wavelength tensor mode is same as  radiation fluid,
and the long wavelength tensor mode is same as  positive spatial
curvature. This result is consistent with the analysis of
Abramo\cite{abramo3} using the effective energy momentum tensor.
\subsection{gauge invariant zero mode solution in the longitudinal
  gauge}

To compare the result in the comoving gauge  with other
gauge condition, we use the longitudinal gauge $E=B=0$ here. We
present only the result of the calculation. The first order 
solution for the scalar mode is
\begin{equation}
  \overset{(1)}g_{ab}=
   \begin{pmatrix}
    -\frac{20}{9}\,\Psi(\boldsymbol{x}) & 0 \\
    0 & -a^2\,\frac{20}{9}\,\Psi(\boldsymbol{x})\,\del_{ij}
   \end{pmatrix},
\end{equation}
and the second order zero mode solution is
\begin{equation}
  \langle\overset{(2)}g_{ab}\rangle =
   \begin{pmatrix}
    -\frac{50}{7\cdot 9^2}\,t^{2/3}\sum_{\bk}k^2\Psi_{\bk}\Psi_{\bk}^{*} &
    0 \\
    0 & -a^2\frac{50}{7\cdot 9^2}\,t^{2/3}\sum_{\bk}k^2\Psi_{\bk}\Psi_{\bk}^{*}
   \end{pmatrix}.
\end{equation}
In this gauge, each components of the first order metric  are gauge invariant
quantities and we can set $\boldsymbol{X}=0$ to construct the second
order gauge invariant quantities. Therefore the gauge invariant zero mode
solution is given by very simple form:
\begin{equation}
  \overset{(0)}Q_{ij}+\overset{(2)}Q_{ij}=\overset{(0)}{g}_{ij}+\langle\overset{(2)}{g}_{ij}\rangle.
\end{equation}
This means the metric components of the line element
\begin{equation}
  ds_{\text{long}}^2=-\left(1+\frac{50}{7\cdot
      9^2}\,t^{2/3}\sum_{\bk}k^2\,\Psi_{\bk}\Psi_{\bk}^{*}\right)dt^2+a^2
     \left(1-\frac{50}{7\cdot
         9^2}\,t^{2/3}\sum_{\bk}k^2\,\Psi_{\bk}\Psi_{\bk}^{*}
     \right)d\boldsymbol{x}^2
\end{equation}
 are  gauge invariant under the first order gauge transformation. As we have the
freedom of the second order coordinate transformation, which is the
second order coordinate transformation of time, we can transform this 
metric to the FRW form. We change the time
coordinate as
\begin{equation}
    t=T-\int dt\,\frac{25}{7\cdot 9^2}\,t^{2/3}\sum_{\bk} k^2\,\Psi_{\bk}\Psi^{*}_{\bk}.
\end{equation}
Then the metric becomes
\begin{equation}        
    ds^{2}_{\text{long}}=-dT^{2}+T^{4/3}\left(a_{0}^{2}
     -\frac{10}{81}\,T^{2/3}\sum_{\bk}k^{2}\,\Psi_{\bk}\Psi^{*}_{\bk}\right)
     d\boldsymbol{x}^{2}.
\end{equation}
At this stage, we use the renormalization method and the second order
perturbation term is absorbed to the constant $a_{0}$ of the zeroth
order. The renormalization group equation is
\begin{equation}
    \frac{\pa \,a^2_{0}}{\pa \,
    T^{2/3}}=-\frac{10}{81}\sum_{\bk}k^{2}\,\Psi_{\bk}\Psi^{*}_{\bk},
\end{equation}
and the solution is
\begin{equation}  
  a_{0}(T)=\left(c-\frac{10}{81}\,T^{2/3}
     \sum_{\bk} k^2\,\Psi_{\bk}\Psi^{*}_{\bk}\right)^{1/2}.  
\end{equation}
The renormalized metric is given by
\begin{align}   
    ds^{2}_{\text{long}}&=-dT^{2}+\left(a^{R}(T)\right)^{2}d
      \boldsymbol{x}^{2}, \notag \\
      &;\quad a^{R}(T)=T^{2/3}\,a_{0}(T).
\end{align}
This metric give the same result in the comoving gauge. Therefore we
have confirmed the gauge invariance of the back reaction effect.

\section{summary and discussion}

We have investigated the cosmological back reaction problem using the
second order perturbation of the Einstein's equation. To describe the
back reaction effect due to the inhomogeneity in the framework of the
cosmological perturbation approach, we used the renormalization group
method. The second order zero mode solution which appeared by the
non-linear effect is regarded as a secular term of the perturbative
expansion, we renormalized a constant of integration contained in
the background solution and absorbed the secular term to this
constant. The renormalized constant obeys the renormalization group
equation which determines the dynamics of the effective background
universe. 

Owing to the gauge freedom of the Einstein's equation, it is
not obvious that what combination of the second order  variables should
be renormalized. Constants contained in the background solution change if we use a $N(t)\neq 1$ lapse function for the background metric. The gauge independent background constants can be obtained by using the Hamilton-Jacobi method\cite{long} and we can recognize the constant $a_0$, which is the freedom of rescaling of the scale factor, has the gauge invariant meaning.  For perturbation, using the second order gauge invariant variable, we
renormalized the secular term to the background constant in a gauge invariant way. In comoving
synchronous gauge, the gauge invariant combination corresponds to a 
comoving volume of the universe. This means that the requirement of
the gauge invariance for inhomogeneous spacetime uniquely picks
out the effective background FRW universe. In previous works, the
effective scale factor was introduced through the uniform Hubble
slicing condition\cite{futamase2,boersma} and the comoving volume
 was introduced from the first\cite{russ2}. Anyway gauge
invariance of the back reaction was not manifest.

For the dust dominated universe, we derived the renormalization group
equation and its solution is obtained. The scalar mode and the long
wavelength tensor mode inhomogeneity work as  positive spatial
curvature, and the short wavelength tensor mode works as  radiation
fluid. These results are consistent with previous
works\cite{russ2,boersma,abramo3}. We also confirmed that these
results are gauge independent by comparing the calculation using the
comoving and the longitudinal gauge.
Besides determining the dynamics of the effective background FRW
universe, the renormalization group method gives the evolution of the
fluctuation which includes the back reaction effect. As the background
evolution is affected by the inhomogeneity, the evolution of the
fluctuation is also modified. For the first order perturbation
solution of the density contrast (\ref{eq:density}), we can obtain the
renormalized evolution by replacing $a_0$ contained in the solution with 
the renormalized value $a_0(t)$. For the scalar mode perturbation, we
have
 \begin{equation}
   \frac{\overset{(1)}\rho}{\overset{(0)}\rho}=-\frac{t^{2/3}}{a_0^2(t)}\nabla^2\Psi
   \propto
   \frac{t^{2/3}}{c-\frac{10}{27}\,t^{2/3}\sum_{\bk}k^2\,\Psi_{\bk}\Psi^{*}_{\bk}}.
 \end{equation}
This indicates that the back reaction effect of the inhomogeneity enhances
the growth rate of the density contrast.

Including the scalar field is an interesting task to investigate the
back reaction problem in the inflationary cosmology. This was partly
done by Abramo {\it et al.}\cite{abramo1,abramo2,mukuhanov1} using the
effective energy momentum tensor, but they do not derived explicit
form of the evolution equation of the effective scale factor and the
scalar field. We will treat this subject in a separate publication.

\acknowledgements{The author would like to thank Y. Y. Yamaguchi for
  discussing on the renormalization group method. This work was
  supported in part by a Grant-In-Aid for Scientific Research of the
  Ministry of Education, Science, Sports and Culture of Japan (11640270).}

\begin{thebibliography}{10}

\bibitem{kodama}
H. Kodama and M. Sasaki, Prog. Theor. Phys. Suppliment {\bf 78},  1  (1984).

\bibitem{mukhanov2}
V.~F. Mukhanov, H.~A. Feldman, and R.~H. Brandenberger, Phys. Rep. {\bf 215},
  203  (1992).

\bibitem{issacson}
R.~A. Issacson, Phys. Rev. {\bf 166},  1263  (1968).

\bibitem{futamase1}
T. Futamase, Mon. Not. R. astr. Soc. {\bf 237},  187  (1989).

\bibitem{futamase2}
T. Futamase, Phys. Rev. D {\bf 53},  681  (1996).

\bibitem{russ2}
H. Russ, M.~H. Soffel, M. Kasai, and G. B{\"o}rner, Phys. Rev. D {\bf 56},
  2044  (1997).

\bibitem{boersma}
J.~P. Boersma, Phys. Rev. D {\bf 57},  798  (1998).

\bibitem{abramo1}
L.~R. Abramo, Ph.D. thesis, Brown University, 1997, BROWN-HET-1096,
  gr-qc/9704049.

\bibitem{abramo2}
L.~R. Abramo, R.~H. Brandenberger, and V.~M. Mukhanov, Phys. Rev. D {\bf 56},
  3248  (1997).

\bibitem{mukuhanov1}
V.~M. Mukhanov, L.~R. Abramo, and R.~H. Brandenberger, Phys. Rev. Lett. {\bf
  78},  1624  (1997).

\bibitem{abramo3}
L.~R. Abramo, {Energy Density and Pressure of Long Wavelength Gravitational
  Waves}, preprint astro-ph/9903270, 1999.

\bibitem{chen}
L.-Y. Chen, N. Goldenfeld, and Y. Oono, Phys. Rev. E {\bf 54},  376  (1996).

\bibitem{kunihiro}
T. Kunihiro, Prog. Theor. Phys. {\bf 94},  503  (1995).

\bibitem{sasa}
S. Sasa, Physica D {\bf 108},  45  (1997).

\bibitem{nozaki}
S. Goto, Y. Masutomi, and K. Nozaki, Prog. Theor. Phys. {\bf 102},  471
  (1999).

\bibitem{nambu}
Y. Nambu and Y. Y. Yamaguchi, Phys. Rev. D {\bf 60},  104011  (1999).

\bibitem{boya1}
D. Boyanovsky, H. de Vega, R. Holman, and M. Simionato, Phys. Rev. D {\bf 60}, 065003 (1999).

\bibitem{boya2}
D. Boyanovsky, H. de Vega, and S. Y. Wang, Phys. Rev. D {\bf 61}, 065006 (2000).

\bibitem{ei}
S. I. Ei, K. Fujii, and T. Kunihiro, Annals Phys. {\bf 280}, 236 (2000).

\bibitem{tomita}
K. Tomita, Prog. Theor. Phys. {\bf 37},  831  (1967).

\bibitem{kasai}
M. Kasai, Phys. Rev. D {\bf 52},  5605  (1995).

\bibitem{long}
Y. Nambu and A. Taruya, Class. Quantum Grav. {\bf 15},  2761  (1998).

\end{thebibliography}

\end{document}